# Fibonacci sequence of twist angles in superconducting multi-layer graphene and hydrogenated graphitic fibers

Nadina Gheorghiu[1,4], Charles R. Ebbing[2], George Y. Panasyuk[3],
and Timothy J. Haugan[4]

[1] Previously with UES, Inc., Dayton, Ohio 45432
[2] University of Dayton Research Institute, Dayton, Ohio 45469
[3] National Research Council, Research Associateship Programs, Washington DC 20001, USA
[4] The U.S. Air Force Research Laboratory (AFRL), Aerospace Systems Directorate,
AFRL/RQQM, Wright-Patterson AFB, Ohio

[*]E-mail: Nadina.Gheorghiu@yahoo.com

**Abstract.** A range of twist angles between adjacent surfaces/volumes are intrinsic to natural graphite or artificially design in multi-layer graphene. In addition, stacking faults can be created by the application of mechanic, electric or magnetic fields. Charge and spin transport then occur in relation to the existing twist-angle pattern. In two dimensions, a saddle point in the electronic band structure leads to divergence in the density of states, known as van Hove singularities (vHs). The energy difference between vHs for the conduction and valence bands was found to increase with the twist angle $\theta$ between neighboring graphite domains with respect to the $c$ axis (perpendicular to the graphite planes). In this work, we estimate $\theta$ for the superconducting (SC)-like nano-size multi-layer granular domain in hydrogenated graphitic fibers [1]. We show that this value for $\theta$ and the values found by others for few-layer graphene might actually form the Fibonacci mathematical sequence. Moreover, SC hydrogenated graphite can harbour higher-order topology as reflected in at least quadratic energy gap flattening. Charge transport and magnetization measurements on hydrogenated graphitic fibers have been done using a Quantum Design Physical Properties Measurement System.

**1. Introduction**

Nature is an amazing display of patterns that can be rigorously described by the laws and principles of physics and mathematics [2]. A very distinct geometrical motif was revealed in the 13th century Italy by Fibonacci as the sequence of numbers 0, 1, 1, 2, 3, 5, 8, 13, 21, 34, 55, 89... obtained from the recurrence relation $f_{n+2} = f_{n+1} + f_n$. The Fibonacci sequence describes the pattern of tree branches, pine and pineapple fruits, chamomile flower seeds (Fig. 1, left), the inflorescence of artichokes and sunflowers, honeybees, as well as the man-made resistor ladder or market trading. One significant archaeological discovery found that the dimensions chosen for the creation of a five-to-seven-thousand-year old Cucuteni collection (Fig. 1, right) in today's Romania [3] form a Fibonacci sequence.

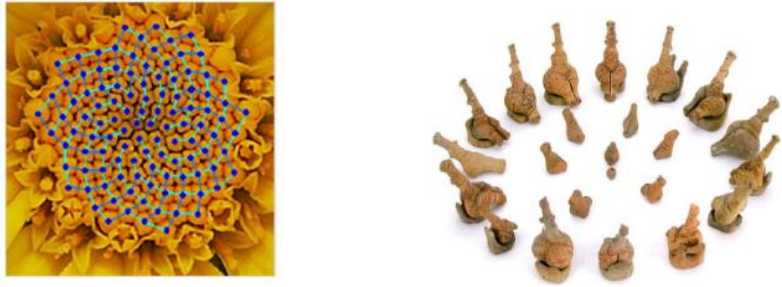

**Figure 1.** Chamomile head showing the spiral pattern of consecutive Fibonacci numbers (left). The 'Council of the Goddesses' from the Isaiia site in today's Romania includes: 13 thrones, 21 female statuettes, 21 cult objects, and 42 beads/caryopes (right).

The 'Holy Family' grouping, also from Isaiia, includes seven anthropomorphic statuettes, plus one throne, thus a total equal to the Fibonacci number 8. The *golden ratio*, which ubiquitous presence in nature was first noticed by Greeks, was later defined by Kepler as the limit ratio $\lim_{n\to\infty} \frac{f_{n+1}}{f_n} \approx$ 1.61803 ... In Egypt, Fibonacci recognized the same mathematical order in the build-up of pyramids [4]. The bioinspired Fibonacci spiral is being implemented in new technological projects. The Fibonacci chain (a quasicrystal), is host to a rich multi-fractal spectrum of SC-normal-SC and SC-insulator-SC states with different topological gaps. The ballistic Josephson direct current established through the weak links of a Fibonacci chain attached to two SC leads probes topological invariance and high $T_c$ [5]. The phonon dynamics in finite Fibonacci superlattices, where layers of two materials (metallic, semiconductor, SC, dielectric, ferroelectric, ceramics) are arranged accordingly to the Fibonacci sequence, have been investigated [6]. The Fibonacci-like sequences reveal hidden symmetries within life's amino acids [11]. In water, the hydrogen (H) atoms covalently bound to oxygen (O) atom such that both bond lengths and bond angles are related to icosahedral symmetry element $\Phi = \frac{1}{2}(1+\sqrt{5}) =$ 1.61803 ... Moreover, the organization of water molecules in chains, clusters, and multi-shells possesses icosahedral symmetry, with solution for Φ forming Fibonacci sequences: 3/2 =1.500, 5/3 = 1.667, 8/5 = 1.600, 13/8 = 1.625, 21/13 = 1.616, 34/21 = 1.619, 55/34 = 1.618, 89/55 = 1.618, 144/89 = 1.618,… [12]. As the human body is about 60% water, the H bonds are clearly essential for life. The H bonding is central to biomolecular functionality by causing protein folding or misfolding, i.e., normal and abnormal protein behavior. A hydrogenated nanomaterial such as fullerenes ($C_{60}$) in water can have a harmonizing role, thus beneficial to proteins. Surely this has to do with the harmony of the genetic code [7]. The four important biomolecules - DNA, microtubules, clathrin and collagen - are structured by Fibonacci sequences and the Golden Mean (*GM*). The basic *GM* ratio results from the division of a unit segment line AB into two parts: first $x$ and second $1-x$, such that $\frac{x}{1-x} = \frac{1}{x}$, with the physical solutions for $GM = \left|\frac{-1\pm\sqrt{5}}{2}\right| = 1.618 \ and \ 0.618$. The generalized *GM* is a solution of the Euler's "divine equation" $\frac{(a+b^n)}{n} = x$. A rectangle in which the side ratio $a/b$ is *GM* can be repeatedly applied and nested to infinity, resulting in a logarithmic spiral often found in nature (nautilus, snail shells, Milky Way). The presence of Fibonacci patterns in organic matter - mainly made of carbon (C), H, and O atoms – inspires important research topics in the fundamental areas of physics, mathematics, chemistry, and engineering. The *GM* ratios are also found in the linear proportions of normal human, animal, and plant bodies, or architectural masterpieces (Greek, Roman), painting and sculptures (Da Vinci, Dali). A *GM* defined as the solution to $x^n + x = 1 \ (n = 1,2,3,...)$ is used in cryptography [8].

    Graphite is a semimetal. Its electronic energy spectrum has topologically protected point nodes. The

point nodes in each layer transform to the chain of the electron and hole Fermi surfaces, which corresponds to approximate line of zeroes protected by topology. The topic of higher-order band topology in twisted multi-layer systems has been recently reviewed, though no connection to the natural world was there mentioned [9]. A two-dimensional (2D) second-order topological phase has gapped 1D edge states and 0D corner states within the energy gap. A 3D second-order topological phase has gapped 2D surface states and gapless 1D hinge states, while a 3D third-order topological phase has gapped surface and hinge states, as well as in-gap corner states. The three-dimensional (3D) or bulk graphite has topologically protected flat band (FB) energy spectrum [10]. The quasiparticle excitation energy at momentum $\boldsymbol{p}$ is $E_{\boldsymbol{p}}(\Delta) = \sqrt{\varepsilon_{\boldsymbol{p}}^2 + \Delta^2}$, where the normal-state dispersion for $s$-wave SC in conventional metals is given by $\varepsilon_{\boldsymbol{p}} = v_F(p - p_F)$, with $v_F$ and $p_F$ the quasiparticle's velocity and the momentum at the isotropic Fermi surface. Then the energy gap is $\Delta = \varepsilon_{\text{uv}}\exp[-1/(|g|v_F)]$, with $\varepsilon_{\text{uv}} \ll \varepsilon_F$ the ultraviolet cut-off energy, such as the one determined by the Debye temperature, $k_B T_D$, where $k_B \cong 1.38 \times 10^{-23}$ J/K is the Boltzmann constant. A totally different situation is when the normal state has a FB, i.e., a region in the momentum space $\boldsymbol{p}$ where $\varepsilon_{\boldsymbol{p}}$. Then $E_{\boldsymbol{p}}(\Delta) = \Delta \sim |g|$, where $g$ is the interaction strength. This will be later discussed in more detail. In graphite, the FB feature comes from a misfit dislocation array, which is spontaneously formed at the interface between two crystals due to the lattice mismatch. At the interface between two domains with different orientation of their crystal axes, a lattice of screw dislocations emerges. The topological origin of this FB can be also understood in terms of the pseudo-magnetic field created when the lattice relaxes from the strain. The width of the FB decreases with the number of layers. The singular density of state (DOS) appearing on the surface gives rise to 2D (surface) superconductivity. The height of the peak in the voltage-dependent nonlocal differential conductance $G_{diff}(V) = dI(V)/dV$ is proportional to the nonlocal DOS, while the energy gap, defined as the width of $G_{diff}(V)$ at half-height, is a measure of electron-electron ($e - e$) correlations between electrons potentially leading to the formation of Cooper pairs. We have previously found from the $G_{diff}(V)$ experimental data clear evidence of topological phenomena such as interference of chiral asymmetric Andreev edge states and crossed Andreev conversion [1]. $G_{diff}(V)$ has a negative part that results from the nonlocal coherence between electrons and holes in the Andreev edge states. Chirality can induce giant charge rectification in a SC. As we have shown, hydrogenated graphite bears the marks of an unconventional high-temperature superconductor (HTSC). It is the emerged Andreev flat SC band that gives rise to a significant zero-energy peak in the singular DOS.

In our earlier work [1], we have found that the temperature-dependence of the SC gap for hydrogenated graphitic fibers, which was extracted from nonlocal differential conductance data, followed the flat-band energy relationship. When a magnetic field was applied normal to the fiber's length, both the electrical resistivity and the magnetization loops showed ferromagnetic (FM) features. The system was found to resemble an unconventional spin-triplet FM SC (or FMSC), in which the $e - e$ correlations are mediated by magnetic correlations introduced by hydrogenation with an alkane. Hydrogenation was found to effectively trigger magnetic correlations in graphite/graphitic systems [10]. Coincidentally, a nearly flat energy band can be spontaneously formed as a result of a misfit dislocations at graphite's interfaces. Mechanical strain on a graphitic system is equivalent to a pseudo-magnetic field [13] and results in FM behavior. The strain can also lead to relative twist in the alignment of neighboring interfaces, resulting in the formation of stacking faults between regions of different twist angles with respect to the common $c$-axis or between regions with different stacking orders [14]. It is as if the dynamic equilibrium of the system, as shown in its charge and spin transport patterns, is achieved for a particular set of twist angles between neighboring grains. One can think of the dynamic equilibrium as being determined by the interplay between the two length scales that are characteristic to the two kinds of order: the smaller periodic crystalline order with the phonons as the major players and the longer quasiperiodic order that is determined by the sequential deposition of different layers and where the $e - e$ correlations can play the major role. At their structural level, graphitic fibers are turbostratic, with volumes of parallel nearest-neighbor graphitic layers randomly rotated such that the overall structure is a random collection of 2D graphitic domains on the smaller scale while quasi-1D on the larger scale. Small angle X-ray scattering shows that the C fibers are fractal objects, with their mass scaling relationship given by $M = L^{d_H}$, where $d_H$ is the Hausdorff dimension. The scattering intensity for the polyacrylonitrile ((CH2-CH-CN)n) or PAN-derived graphitic fibers varies as $I(S) \propto S^{2/3}$, where and $S = \Delta k/2\pi$ takes values between 1.5 and 3 nm$^{-1}$, with $k$ being the wavevector. The disordered nature of these C fibers was reviewed in [15]. The competition between disorder and $e$ - $e$ correlations is an important topic in condensed matter physics, though the effect of disorder on SC is yet to be understood. The effect of Fibonacci-modulated disorder via a Fibonacci modulated sequence of the site potentials on two different models of SC supporting short-range pairing was studied [16]. There is increased backscattering in the presence of Fibonacci disorder. In [17], we have found proof of ballistic 1D transport as a minimum in the nonlocal differential conductance that is known to be due to Zeeman spin-splitting. The conductance peak in high enough magnetic field ($B \sim 1$ T) is a signature of ballistic 1D transport, where backscattering is reduced and spin degeneracy is lifted [18].

A tilt grain boundary on a top graphene sheet in multi-layer graphene leads to a twist between consecutive layers and generates superstructures (i.e., Moiré patterns) on one side of the boundary. This results in changed electronic properties of graphene bilayers. The system on one side of the grain boundary is changed from Bernal to twisted graphene bilayer. The parabolic energy spectrum of Bernal graphene bilayer is split into two Dirac cones. Thus, for the twisted graphene bilayer, two saddle points along the intersections of two Dirac cones appear in the low-energy band spectrum. The saddle points result in two vHs in the DOS. The energy difference of the two vHs is linear in the twist angle. Tuning the chemical potential at a vHs introduces a large number of single-particle states with negligible energy likely to form correlated states. In particular, vHs in mirror-symmetric twisted trilayer graphene was found to be associated with a strong zero-energy peak in the DOS. In contrast with twisted bilayer graphene where SC was found at a magic angle $\theta \cong 1^0$ [19], vHs arise from the fusion between the standard finite energy of the vHs and the Dirac cone at point K of the Brillouin zone, where the π-symmetry bands linearly intersect [20]. It appears that more than two-layer magic-angle graphene structures show increasingly more opportunities for $e$ - $e$ SC correlations leading to higher critical temperatures $T_c$.

In this paper, we estimate a twist angle for the SC nano-size multi-layer granular domains in hydrogenated graphitic fibers to $\theta_{H-C} \cong 13^0$. In addition, we show that the twist angles found by others for few-layer graphene [21] and our estimate for $\theta$ might actually form a Fibonacci sequence. We interpret this finding as yet another confirmation that SC in hydrogenated graphite has topological manifestations. It is an open question whether the Fibonacci sequence of twist angles results from the Fibonacci-modulated sequence of site potentials leading to vHs. As we will see, our analysis suggests such a possibility.

## 2. Experiment

The materials used for the twist angle analysis are 7 μm in diameter polyacrylonitrile ((CH$_2$–CH–CN)$_n$) or PAN-based fibers of T300 type with a C content ∼ 93%. Hydrogenation is achieved by the use of an alkane (octane, C$_8$H$_{18}$). Resistivity measurements on the hydrogenated fibers were conducted with the temperature ($T$)-stabilized by the surrounding cryogenic fluid in a 6500 Quantum Design of the Physical Properties Measurement System (PPMS) [22]. Direct current electrical currents were sourced through a Keithley 2430 1 kW PULSE current-source meter and voltages for gaps (i.e., wire/fiber lengths) up to 2 mm along the fibers were measured with a Keithley 2183A Nanovoltmeter. More experimental details can be found in [1].

## 3. Results and Interpretation

In [1], we have discussed the case of HTSC in hydrogenated graphitic fibers on the basis of Little model for HTSC in organic materials [23]. The molecular arrangement has two parts: a) a long chain called spine in which the electrons fill various states and may or may not form a conducting system; b) a series of arms or side chains attached to the spine. Charge and spin density waves in the side chains are established resulting in $e-e$ correlations, some of which are SC. Thus, the spine transforms from the insulating or semiconducting state directly to the SC metallic state upon the addition of the side chains. In the case of hydrogenated graphitic fibers or graphite foils [24], the treatment with octane can also result in freely moving protons (H$^+$) on the PAN fibers' or graphite's surfaces [25]. The octane $C_8H_{18}$ molecule itself has a spine formed by its eight C atoms and the side chains are its eighteen H atoms. As the length of the C–C and C–H bonds are 1.54 Å and 1.10 Å, respectively, the octane molecule is approximately $L = 1.3$ nm in length and $w = 2.20$ Å wide. Interestingly, perhaps not surprisingly in the framework of Little model, the $e-e$ Cooper pairing length is the length of the octane molecule, $\xi_0 \cong L$. Moreover, from Little-Parks oscillations at 300 K, we have found $\xi_{LP} \cong 1.4$ nm [1]. Notice also that $\xi_0$ is close to the lower limit for the scattering intensity dependence $I(S) \sim S^{2.3}$, where $S = (1.5 \div 3)$ nm$^{-1}$. As known, the critical field for a type II SC is given by: $B_{c2} = \frac{\Phi_0}{2\pi \xi_{GL}^2}$, where $\Phi_0 = h/2e \cong 2.07 \times 10^{-15}$ T·m$^2$ is the magnetic flux quanta, with $h = 6.62 \times 10^{-34}$ J·s the Planck constant and $e = 1.9 \times 10^{-19}$ C the electron's charge. From the magnetization loop of an octane-graphite powder mixture at 60 K (Fig. 2), we found $B_{c2} \cong 50$ mT. Then the Ginzburg-Landau coherence length is $\xi_{GL} \cong 82$ nm. Since both the C fiber and the octane-powder mixture show the low-$T$ critical transition for $T \sim 50$ K, we can use the $\xi_{GL}$ just found to estimate the coherence length for the case of a type II dirty SC. The Ginzburg-Landau formula gives: $\xi_{GL}^{dirty} = 0.85(l\xi_0)^{1/2}(1 - T/T_c)^{-1/2}$, where $l$ is the mean-free path for electrons between two elastic collisions. Previously, we have estimated from all $T$-dependent gap $\Delta(T)$ data that there could be a SC phase with $T_c$ as high as 440 K as shown (Fig. 3, from [1]). We then find $\xi_{GL}^{dirty}(T = 0\ K) = \left(\frac{\xi_{GL}^{dirty}(T=60\ K)}{0.85}\right)^2 \times \left(1 - \frac{60\ K}{440\ K}\right) \times \frac{1}{120\ nm} \cong 76$ nm and $\xi_{GL}^{dirty}(T = 300\ K) = 0.85\sqrt{\frac{(67\ nm)(120\ nm)}{1-300/440}} \cong 135$ nm. This is about two times the radius of the fluxoid for the Little-Parks oscillations. Thus: $\xi_0 / \xi_{GL}^{dirty}(T = 0\ K) \cong 0.020 = 2\%$. Since: $\frac{\xi_0}{\xi_{GL}^{dirty}(T=0\ K)} \cong \frac{\Delta_c}{\Delta_p} = \frac{T_c}{T_p}$, then $T_p \cong \frac{67\ nm}{1.3\ nm} T_c \cong 2.3 \times 10^4$ K, where $\Delta_{c,p}$ and $T_{c,p}$ are the coherence/pairing gap and the critical/pairing temperatures, respectively. Within the Bardeen-Cooper-Schrieffer (BCS) theory [26], $T_c$ is determined by the linearized gap equation. Schrieffer found that the spin deformation potential leads to $p$-wave ($l = 1, s = 1$) pairing for ferromagnetic FM spin fluctuations, at a remarkably high temperature of order $T_c \cong 3.0 \times 10^4$ K [27]. This could be the temperature at which bisolitons (pairs of polaronic solitons) are formed. As known and also found here, the Cooper pair length $\xi_0$ and the phase coherence length $\xi_{GL}$, are two different lengths. This is because SC requires two different physical phenomena: the electron pairing and the onset of long-range phase coherence. The coherence length $\xi_{GL}$ is the length over which variations of the order parameter of the superconducting condensate occur, whilst the pair size $\xi_0$ is related to the wavefunction of a Cooper pair. While the coherence length depends on $T$, $\xi_{GL}(T)$, the Cooper-pair size is $T$-independent. The coherence length diverges at $T \to T_c$ (Fig. 3, right), suggesting that the hydrogenated graphite fiber is a heavy fermion system.

Further, application of the Little model to the hydrogenated graphitic fiber suggests that the length of the octane molecule is also the length of a perfect Bernal graphite domain, $L \cong \xi_0 \cong 1.3$ nm. The angle of rotation with respect to the $c$ axis between well-oriented graphite domains is $\theta_{twist} = 2\sin^{-1}(a/L)$, where $a = 1.42$ Å is the graphite's nearest neighbour C-C distance. Thus, we find that the twist angle in the hydrogenated C fiber is $\theta_{H-C} \cong 13^0$. Notice that a twist angle $\theta_{twist}/2 = 30^0$ gives $L = \frac{a}{\sin(30^0)} \cong 2.84$ Å. In this case, the perfect graphite domain is the hexagonal prism unit and the twisting changed nothing, i.e., the graphite symmetry is rhombohedral. As known, rhombohedral multi-layer graphene or graphite are SC due to its large surface DOS. Any graphite sample contains some fraction of rhombohedral stacking, as we have also found from analysing the X-ray diffraction peaks for a natural graphite sample.

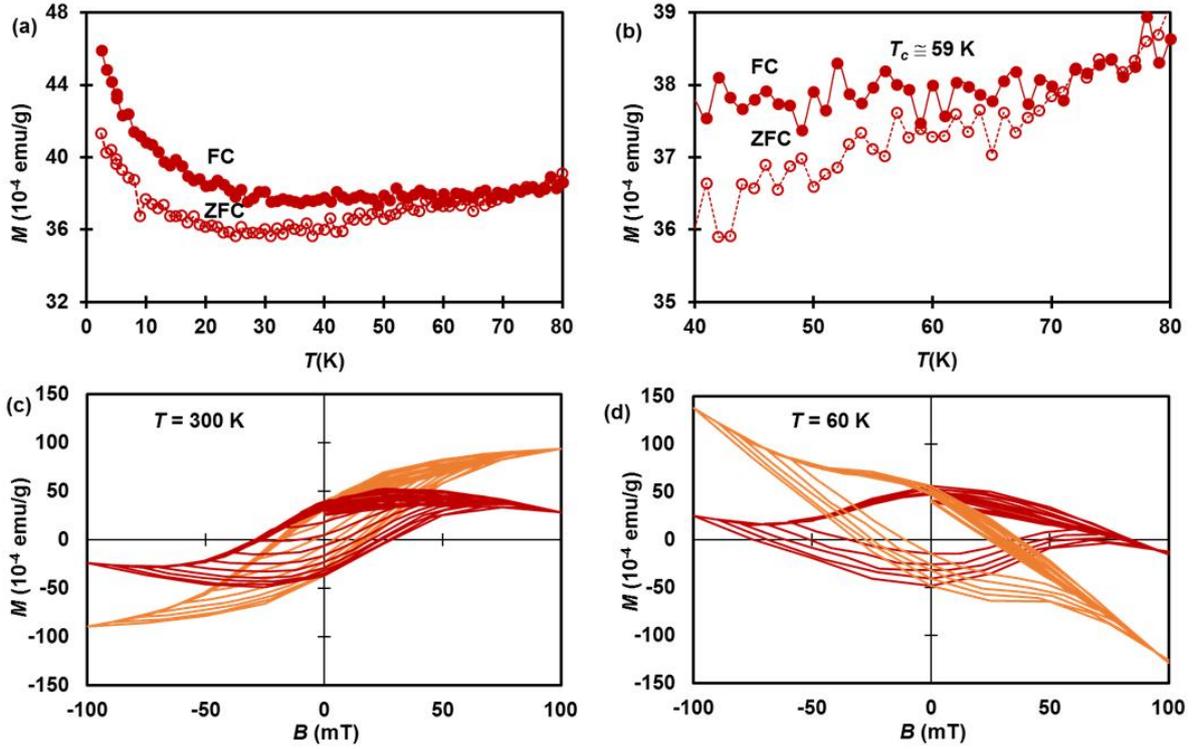

**Figure 2.** Temperature dependent magnetization data (a, b) is positive, reflecting the low-temperature paramagnetic SC. Magnetization loops for octane-graphite powder mixture (c, d) show high-temperature ferromagnetism. Diamagnetic data included (in orange) and excluded (in red).

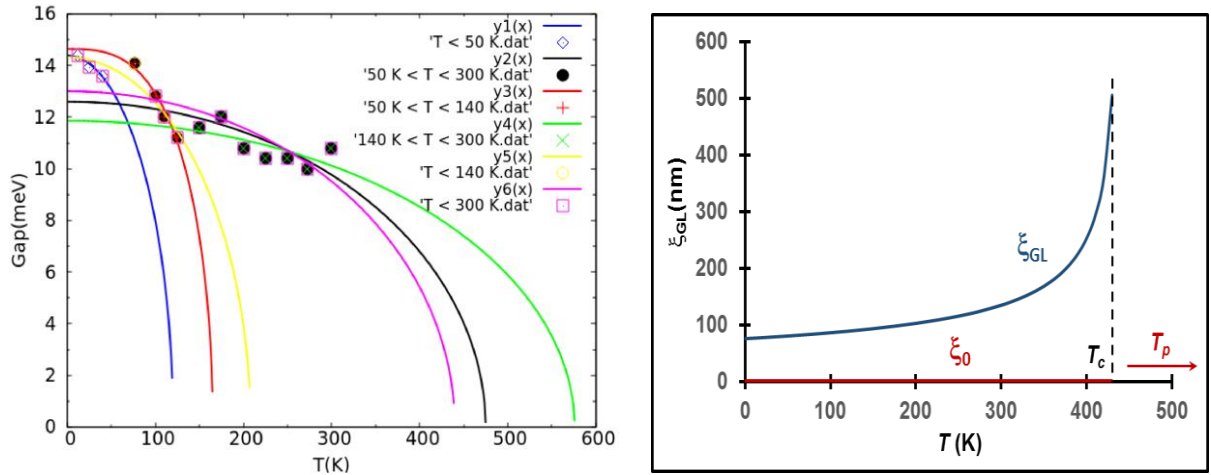

**Figure 3.** Temperature dependence of the energy gap obtained from nonlocal differential conductance measurements (left). The temperature dependence of the phase coherence length (right).

The main topic is now in sight. The twist angle we have just estimated, $\theta_{H-C} \cong 13^0$, is a Fibonacci term when starting with the close to the $\theta \sim 1^0$ value found for bilayer graphene, $\theta \cong 1.08^0$ [19]. Other twist angles from the Fibonacci sequence, such as $\theta \cong 21^0$ (close to the $22^0$ twist angle corresponding to graphene's lattice constant 2.46 Å) and $\theta \cong 89^0$ (the strained and twisted vector fields are related to each other by a $90^0$ rotation angle), or twist angles described by the golden ratio [28], can also lead to high DOS and SC. It was also found that a tilt grain boundary on a top graphene sheet in graphene multi-layers leads to a twist between consecutive layers and generates superstructures (i.e., Moiré patterns) on one side of the boundary. This results in changed electronic properties of graphene bilayers. The system on one side of the grain boundary is changed from Bernal to twisted graphene bilayer. The parabolic energy spectrum of Bernal graphene bilayer is split into two Dirac cones. Thus, for the twisted graphene bilayer, two saddle points along the intersections of two Dirac cones appear in the low-energy band spectrum. These saddle points result in two vHs in the DOS. The energy difference between vHs states from the conduction and valence bands was found to depend linearly on $\theta$ (with respect to the $c$ axis) between neighboring graphite domains: $\Delta E_{vHs}(\theta) = a\theta + b \cong 0.18\theta - 0.20$ (eV), where the coefficients are approximate values taken from Fig. 2b in Ref. [21]. Using the value for $\theta_{GL} \cong 13^0$ that we have found before for the hydrogenated graphitic fiber, we extract the vHs energy gap: $\Delta E_{vHs}(\theta) \cong$ 2.1 eV. This value is close to the intensity enhancement of the G mode for twisted tetralayer graphenes as determined by Raman spectroscopy. Using the data for twisted bilayer graphene, $\theta_{twist} \cong 1.08^0$ [19], $\Delta E_{vHs} \cong 12$ meV [19], and $T_c \cong 1.7$ K, the critical temperature for the hydrogenated graphitic fiber is estimated to $T_c \cong \frac{2.1 \text{ eV}}{0.012 \text{ eV}} \times 1.7 \text{ K} \cong 300$ K. This is a HTSC, where the high DOS was achieved by the hydrogenation of its many interfaces.

Moreover, in our previous work we have found from the energy gap $\Delta(T)$ data that there could be a SC phase with $T_c$ as high as 440 K. The latter is a higher value than the one inferred from the twist-angle dependence of the vHs energy. If the hydrogenated graphite would be a conventional SC with linear dispersion, then $T_c = T_D e^{-1/g}$ [26]. Here, $g = N(0)V$ is the interaction strength in the Cooper channel, $N(0)$ is the DOS near the Fermi surface in the normal state, and $V$ is the matrix element of the attractive potential between particles. Various estimates give $g \sim 0.1 - 0.5$ [29]. For $g \cong 0.5$ and using the known Debye temperature $T_D \cong 2430$ K for graphite, we find $T_c \cong 324$ K. The higher $T_c \cong 440$ K would correspond to $g \cong 0.25$. Yet, as our previous work already showed, hydrogenated graphite has a FB energy system, hence it is not a conventional HTSC. The existence of FBs leads to singular DOS, $\frac{dN(E)}{dE} \to 0$ as $E \to 0$. It results in $T_c \propto g$ for 2D systems [30]. In 1D systems, the quadratic flattening gives $N(E) \propto E^{-1/2}$, resulting in the quadratic dependence of $T_c$ on the interaction strength, $T_c \propto g^2$ [31, 32]. While the vHs result from the non-topological $e$ - $e$ correlations, the *higher-order topology* of the energy gap flattening suggests that the hydrogenated graphitic fibers harbour 1D (filamentary) HTSC [33]. This is as if charge carriers flow would occur through 1D topologically protected channels. Notice that $T_c(g) \sim g - g^2$ has the maximum at $g \cong 0.5$. Thus, the high value $T_c \cong 440$ K obtained from the $\Delta(T)$ data [1] can be explained by considering at least the quadratic gap flattening characteristic to an unconventional SC. The high $T_c$ values (Fig. 3) and high excitonic SC gap 1.6 eV that we have found [1], are proofs that Little's proposal has always been within reach [34].

## 3. Conclusion

We propose that the twist angle for the hydrogenated graphitic fiber, $\theta_{H-C} \cong 13^0$, and the twist angles found by others for few-layer graphene might actually form a Fibonacci mathematical sequence. Moreover, we predict that SC might be observed for other twist angles of the Fibonacci sequence. While conceptually daring, this interpretation aligns with our previous findings regarding the topological nature of the flat-energy band SC in hydrogenated graphite [1]. Moreover, we have found that the high-$T_c$ SC for hydrogenated graphite can only be explained by the existence of higher-order topology quantified as at least quadratic energy gap flattening.

## Acknowledgments

This work was supported by The Air Force Office of Scientific Research (AFOSR) and the Aerospace Systems Directorate (AFRL/RQ) through the LRIR grant #18RQCOR100. We acknowledge early discussions with Dr. T. J. Bullard, as well as technical assistance from J.P. Murphy and J. Lawson.

## References

[1] Gheorghiu N, Ebbing CR, and Haugan TJ 2020 *arXiv:2005.05876*
[2] Adam JA 2003 *The Fibonacci Sequence and the Golden Ratio*, in book *Modeling Patterns in the Natural World*. (Princeton University Press)
[3] Ursulescu N and Merlan V 1997 *Isaiia – Balta Popii* 32 (Cronica Cercetărilor Arheologice din România)
[4] Pesavento L 1997 *Fibonacci Ratios with Pattern Recognition* (Traders Press Incorporated)
[5] Sandberg A *et al*. *Phys Rev B* 2024 **110** 104513
[6] Maciá E 2006 *Phys Rev B* **73** 184303
[7] Rakočević MM 2017 *Harmony of the genetic code* vols 1&2 (online)
[8] Stakhov AP 1989 *Computers Math Applic* **17** 613
[9] Hua C and Xu D-H 2025 *Chin Phys B* **34** 037301
[10] Esquinazi P ed 2016 *Basic Physics of Functionalized Graphite* (Springer)
[11] Négadi T 2024 *Symmetry* **16** 293
[12] Matija LR *et al* 2023 *Micromachines* **14** 2152
[13] Volovik GE 2018 *JEPT Letters* **107(8)** 516
[14] Ariskina R, Stiller M, Precker CE, Böhlmann W, and Esquinazi P 2022 *Materials* **15** 3422
[15] Dresselhaus MS, Dresselhaus G, Sugihara K, Spain II, and Goldberg HA 1988 *Graphite Fibers and Filaments* (Springer-Verlag)
[16] Gupta S, Sil S and Bhattacharyya B 2006 *J Phys: Condens Matter* **18** 1987
[17] Gheorghiu N, Ebbing CR, Haugan TJ 2020 *Conf Ser: Mater Sci Eng* **756** 012022
[18] Estrada Saldaña JC *et al*. 2019 *Sci Adv* **5** 1
[19] Cao Y *et al* 2018 *Nature* **556** 45
[20] Guerci D, Simon P, and Mora C 2022 *Phys Rev Research* **4** L012013
[21] Yin L-J *et al* 2014 *Phys Rev B* **89** 205410
[22] Quantum Design USA
[23] Little WA 1964 *Phys Rev A* **134(6A)** 1416
[24] Gheorghiu N, Ebbing CR, Haugan TJ 2024 *IEEE Trans Appl Supercond* **35(1)** 7400205
[25] Kawashima Y & Iwamoto M 2016 *Sci Rep* **6** 28493
[26] J Bardeen J, Cooper LN, and Schrieffer JR 1957 *Phys. Rev.* **108** 1175
[27] Schrieffer JR 2004 *Journal of Superconductivity: Incorporating Novel Magnetism* **17(5)**
[28] Khalaf E, Kruchkov AJ, Tarnopolsky G, and Vishwanath A 2019 *Phys Rev B* **100** 085109
[29] Ginzburg VL 1964 *JETP* **47** 2318
[30] Khodel VA and Shaginyan VR 1990 *JETP Lett* **51** 553
[31] Kopaev Y 1970 *Sov Phys JETP* **31** 544
[32] Esquinazi P, Heikkilä TT, Lysogorskiy YV, Tayurskii DA, and Volovik GE 2014 *JETP Lett* **100(5)** 336
[33] Ginzburg VL 1968 *Contemp Phys* **9(4)** 355
[34] Grant PM 1998 *Phys Today* **51(5)** 17